\documentclass[11pt,twoside]{article}

\usepackage{asp2006}
\usepackage{epsf}
\usepackage{epsfig}
\usepackage{lscape}
\usepackage{mathtext}
\usepackage{amssymb}

\begin{document}

\title{IS \emph{"SPIKE"} A RELIABLE FEATURE IN \textbf{$P_{orb}$} DISTRIBUTION OF \emph{AM HER} STARS?}

\author{Alex Golovin}
\affil{Main Astronomical Observatory of National Academy of
Sciences of Ukraine, Zabolotnogo, 27, Kiev, 03680, Ukraine \\
Kyiv National Taras
Shevchenko University, Physics Faculty, Kyiv, Ukraine \\
golovin.alex@gmail.com}
\author{Mykola Malygin}
\affil{Kyiv National Taras Shevchenko University, Physics Faculty,
Kyiv, Ukraine}
\author{Elena P. Pavlenko}
\affil{Crimean Astrophysical Observatory, Crimea, Ukraine}

\begin{abstract}
Orbital periods in AM Her stars (\emph{polars}) are synchronized
with spin periods of white dwarf by its high magnetic field. Since
the last study of $P_{orb}$ distribution of these systems, the
number of known objects of such type has more than doubled. This
challenged us to compile a new updated catalogue of cataclysmic
variables with highly magnetic white dwarfs (polars) and to study
their $P_{orb}$ distribution. In this paper we also discus if
"spike" is reliable feature in the distribution. ("Spike" is a
concentration of polars in the distribution of their orbital
periods near $P_{orb}$ = 114 min and was previously discussed by
Ritter \& Kolb (1992) and Shahbaz \& Wood (1996).)

\end{abstract}
\section{Introduction}

AM Her stars or Polars - is a subtype of cataclysmic variables,
where binary stellar system contains highly magnetic white dwarf.
The strength of magnetic field allows to control accretion in such
system (mass is transferring directly on the white dwarf pole,
without formation of accretion disc). As a distinct from
intermediate polars, AM Her stars orbital period is synchronized
with white dwarf spin period.

Last study of $P_{orb}$ distribution were made by Shahbaz \& Wood
(1996) in 1996 (just 43 systems were known by that time). They
have noted that discovery of one more AM Her-type star with the
orbital period outside of the spike will decrease its significance
below 99\% level. Since that time the number of known systems has
more then doubled. This incentivised us to compile a new updated
catalogue of cataclysmic variables with highly magnetic white
dwarfs and to study distribution of their orbital periods and to
re-calculate significance of spike in the similar way, as it was
done in previous work.

\section{Results}

Our Catalogue was compiled based on recently published papers on
newly discovered polars and previous catalogues by Katysheva \&
Voloshina (2007)and Ritter \& Kolb (2003). It contains an
information about 91 polars (designation, coordinates, $P_{orb}$,
masses of components (if known), strength of white dwarf magnetic
field).

They have orbital periods ranging from 77 minutes to 286 minutes.
We set period range for "spike" as 113-117 minutes. We obtained a
grater statistic (91 systems in total) and were impelled to make
the spike ranges some broader: $\Delta P_{spike}$ = 4 min ($\Delta
P_{spike}$ = 2 min in Ritter \& Kolb (1992), $\Delta P_{spike}$ =
3 min in Shahbaz \& Wood (1996)). Totaly, 11 AM Her-type stars
have orbital periods "in spike".

\begin{figure}[!h]
\plotone{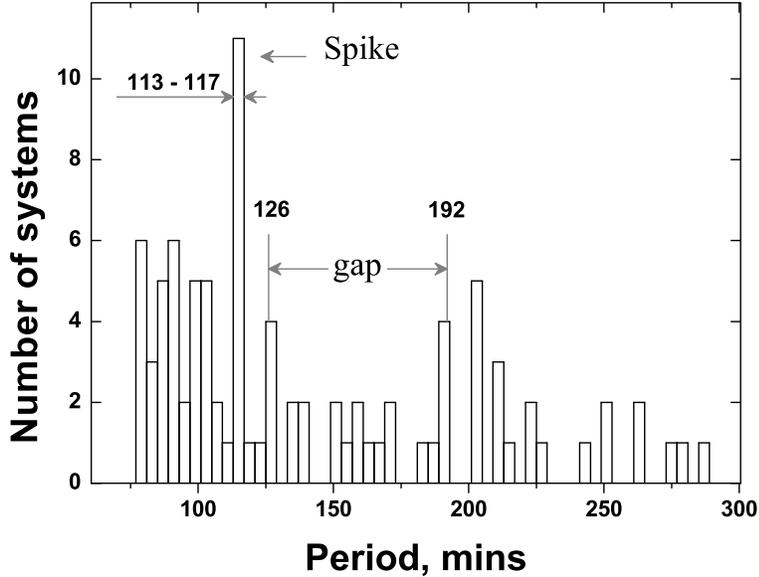}
\caption{$P_{orb}$ Distribution of AM
Her Stars}
\end{figure}

Values of significances were obtained in the same way as in
Shahbaz \& Wood (1996). Spike still appears to be a significant
feature in $P_{orb}$ distribution of polars, although its
significance now is equal to $99.6 \%$  ($99.9 \%$ in previous
work).

Figure 1 represents newly obtained distribution of orbital periods
of highly magnetic cataclysmic variable stars. It could be
concluded that about 55\% of polars are concentrated below "period
gap". Future study of polars with orbital periods in spike are
critically important, since they might have very similar masses of
components. Ritter \& Kolb (1992) have shown, that dispersion of
the secondary masses in the period spike must be
\begin{math} \delta M_2  \lesssim 2 \cdot 10^{-3} M_{\odot}
\end{math}. Masses of white dwarf considered to be $M_1 \gtrsim
0.7 M_{\odot}$ with mass dispersion of $\delta M_1 \lesssim 0.05
M_{\odot}$

The full version of the Catalogue and detailed paper on our
analysis will be published elsewhere.

\section{Conclusion}

\begin{itemize}
    \item Most of known polars have orbital periods below "period
    gap" (50 of 91 known polars).
    \item We obtained a grater statistic (91 systems in total) and were impelled to make the spike ranges some broader: $\Delta P_{spike}$ = 4 min ($\Delta P_{spike}$ = 2 min in Ritter \& Kolb (1992) - 17 polars in total were known, $\Delta P_{spike}$ = 3 min in Shahbaz \& Wood (1996) - 43 polars were
    known).
    \item Though spike remains to be significant feature, its
    significance has decreased and now its value is about $99.6 \%$, instead of  the 99.9\%, as it was in the work of T.Shahbaz \& Janet H.Wood (1996).
    \item The observed number of systems in the gap is inconsistent with the period distribution being uniform at the 96.3\% level.  So we have a lowering of this value, just like as for the spike significance.

\end{itemize}

\acknowledgements This work was partially supported by the grant
of the FRSF  F25.2/139. A. Golovin work was partly supported by
Ekialde Foundation.

\end{document}